\documentstyle[12pt]{article}

\begin{document}
\begin{center}
{\LARGE{Dilatonic, Current-Carrying Cosmic 
Strings}\footnote{Talk presented by M. E. X. 
Guimar\~aes at the COSMO 99, Trieste, It.}} 
\end{center}
\begin{center}
{\large{C. N. Ferreira$^{1}$, M. E. X. 
Guimar\~aes$^{2}$ and J. A. Helay\"el-Neto$^{1}$}} \\
1. Centro Brasileiro de Pesquisas F\'{\i}sicas, Brazil \\
2. Depto. de Matem\'atica, Univ. de Bras\'{\i}lia, Brazil
\end{center}

\vspace{1.5cm}

This talk is presented as follows:

\vspace{1.0cm}

I. Motivation

\vspace{0.5cm}

II. The Model $ U(1) \times U'(1)$ coupled to the Scalar-Tensor Gravity

\vspace{0.5cm}

III. Superconducting Cosmic String Solution in Scalar-Tensor Gravity

\vspace{0.5cm}

IV. Discussions and Conclusion

\newpage

\section*{I. Motivation:}

From the theoretical point of view, theories in which gravity is 
mediated by one or more long-range scalar-fields in addition to 
the usual tensor field present in Einstein's gravity are the most 
natural alternatives to General Relativity. Indeed, most attempts 
to unify gravity with the other interactions predict the existence 
of one (or many) scalar(s) field(s) with gravitational-strength couplings. 

If gravity is scalar-tensorial in nature, there will be direct 
implications for cosmology and experimental tests of the gravitational 
interaction\footnote{Th. Damour, {\em gr-qc/9904057}, Nucl. Phys. B (to appear).}: 
one expects that in the early universe the coupling to matter of the scalar component 
of the gravitational interaction would be of the same order of the 
coupling to matter of the long-range tensor component although in the 
present time the observable total coupling strength is generically 
small\footnote{Th. Damour and K. Nordtverdt, 
Phys. Rev. D {\bf 48} (1993), 3436.}.

On the other hand, topological defects are expected to be formed 
during phase transitions in the early universe. Among them, cosmic strings 
have been widely studied in cosmology in connection with structure formation. 
Recently, Witten\footnote{E. Witten, Nucl. Phys. B {\bf 249} (1985), 557.} 
has shown that in many field theories cosmic strings behaves as 
superconducting tubes and they may carry electric charge. Later, it 
was shown that cosmic strings can be current-carrying without the need of 
any external electromagentic field\footnote{P. Peter, Phys. 
Rev. D {\bf 49} (1994), 5052.}.

In this talk we are interested in studying the implications of a scalar-tensorial 
gravity for the metric of an isolated self-gravitating (bosonic) superconducting 
cosmic string. In what follows, we will introduce the gauge model which describes the 
superconducting vortex and we will derive its metric in the weak-field approximation 
in the scalar-tensor gravity. We end this talk with some discussions on the results 
obtained in this work.

\section*{II. The Model:}

We start by introducing the model with which we will deal throughout this work. 
In the Jordan-Fierz frame is is described by the action:
\begin{equation}
{\cal S} = \frac{1}{16\pi} \int d^4x \sqrt{-\tilde{g}} 
\left[\tilde{\Phi}\tilde{R} - \frac{\omega(\tilde{\Phi})}{\tilde{\Phi}}\tilde{g}^{\mu\nu}
\partial_{\mu}\tilde{\Phi}\partial_{\nu}\tilde{\Phi} \right] 
+ {\cal S}_{m}[\Psi_m, \tilde{g}_{\mu\nu}] , 
\end{equation}
where $\tilde{g}_{\mu\nu}$ is the physical metric in this frame, 
$\tilde{R}$ is the curvature scalar associated to it and 
${\cal S}_m$ denotes the action describing the general matter 
fields $\Psi_m$. These theories are metric, e.g., matter couples 
minimally and universally to $\tilde{g}_{\mu\nu}$ and not to 
$\tilde{\Phi}$. 

In what follows, the superconducting vortex arises from a 
spontaneous breakdown of the symmetries of the gauge model 
$U(1) \times U'(1)$. Therefore, the action ${\cal S}_m$ is 
described by:
\begin{equation}
{\cal S}_m =  \int d^4x \sqrt{-\tilde{g}} \left[ 
\frac{1}{2}D_{\mu}\varphi D^{\mu}\varphi^* + \frac{1}{2}
D_{\mu}\sigma D^{\mu}\sigma^* - \frac{1}{4}H_{\mu\nu}H^{\mu\nu}
- \frac{1}{4}F_{\mu\nu}F^{\mu\nu} - V(\mid\varphi\mid , \mid\sigma\mid) \right]
\end{equation}
with $D_{\mu}\varphi \equiv (\partial_{\mu} + iqC_{\mu})\varphi$, 
$D_{\mu}\sigma \equiv (\partial_{\mu} + ieA_{\mu})\sigma$ and 
$F_{\mu\nu}$ and $H_{\mu\nu}$ are the field-strengths 
associated to the electromagnetic $A_{\mu}$ and gauge $C_{\mu}$ 
fields, respectively. The potential 
\begin{equation}
V(\mid\varphi\mid , \mid\sigma\mid) = \frac{\lambda_{\varphi}}{4}
(\mid\varphi\mid^2 - \eta^2)^2 + f\mid\varphi\mid^2\mid\sigma\mid^2 + \frac{\lambda_{\sigma}}{4}
\mid\sigma\mid^4 - \frac{m^2}{2}\mid\sigma\mid^2  ,
\end{equation}
with positive $\eta , f , \lambda_{\sigma}, \lambda_{\varphi}$ 
parameters, is  ``Higgs inspired" and contains appropriate $\varphi-\sigma$ 
interactions so that there occurs a 
spontaneous symmetry breaking. A vortex configuration arises 
when the $U(1)$ symmetry associated with the $(\varphi, C_{\mu})$-pair is 
spontaneously broken. The superconducting 
feature of this vortex is produced when the pair 
$(\sigma, A_{\mu})$, associated with the other $U'(1)$ symmetry 
of this model, is spontaneously broken in the core of the vortex. 

For many reasons, it is better to work in the so-called Einstein (conformal) 
frame in which the kinematic terms of the tensor and scalar fields do not mix:
\begin{equation}
{\cal S} = \frac{1}{16\pi G} \int d^4x \sqrt{-g} \left[ R - 
2g^{\mu\nu}\partial_{\mu}\phi\partial_{\nu}\phi \right] 
+ {\cal S}_{m}[\Psi_m,\Omega^2(\phi)g_{\mu\nu}] ,
\end{equation}
where $g_{\mu\nu}$ is pure rank-2 tensor in the Einstein frame, 
$R$ is the curvature scalar associated to it and $\Omega(\phi)$ 
is an arbitrary function of the scalar field. Action (2) is 
obtained from (1) by a conformal transformation 
\begin{equation}
\tilde{g}_{\mu\nu} = \Omega^2(\phi)g_{\mu\nu} ,
\end{equation}
and by a redefinition of the quantity $
G\Omega^2(\phi) = \tilde{\Phi}^{-1}$, 
which makes evident the feature that any gravitational 
phenomena will be affected by the variation of the gravitation ``constant" $G$ 
in the scalar-tensorial gravity, and by introducing a new parameter $
\alpha^2 \equiv \left( \frac{\partial \ln \Omega(\phi)}{\partial 
\phi} \right)^2 = [2\omega(\tilde{\Phi}) + 3]^{-1}$, 
which can be interpreted as the (field-dependent) coupling strength between 
matter and the scalar field. 

In the conformal frame the "Einstein's" eqs. read:
\begin{eqnarray}
G_{\mu\nu} & = & 2\partial_{\mu}\phi\partial_{\nu}\phi -
g_{\mu\nu}g^{\alpha\beta}\partial_{\alpha}\phi\partial_{\beta}
\phi + 8\pi G T_{\mu\nu} \nonumber \\
\Box_{g}\phi & = & -4\pi G \alpha(\phi) T .
\end{eqnarray}
The energy-momentum tensor is defined 
as usual 
\begin{equation}
T_{\mu\nu} \equiv \frac{2}{\sqrt{-g}}\frac{\delta {\cal S}_m}
{\delta g_{\mu\nu}} ,
\end{equation}
but in the conformal frame it is no longer conserved 
$\nabla_{\mu}T^{\mu}_{\nu} = \alpha(\phi)T\nabla_{\nu}\phi$.

Due to the symmetries of the configuration we are searching for, we set:
\begin{equation}
ds^2 = e^{2(\gamma -\Psi)}(dt^2 - dr^2) - \beta^2 e^{-2\Psi} 
d\theta^2 - e^{2\Psi}dz^2 ,
\end{equation}
where the metric functions $\gamma,\Psi,$ and $\beta$ are functions of $r$ only. 
In addition,  the metric functions 
satisfy the regularity conditions at the axis of symmetry $r=0$
\begin{equation}
\gamma = 0,  \;\; \Psi =0, \;\; \frac{d\gamma}{dr}=0, \;\; 
\frac{d\Psi}{dr} =0,  \;\; \mbox{and} \;\; \frac{d\beta}{dr} =1 .
\end{equation}

In addition to the metric of the form (8), we choose a particular {\em ansatz} 
for the matter fields in order to contemplate an 
isolated superconducting vortex:
\begin{equation}
\varphi = R(r)e^{i\theta} \;\; \mbox{and} \;\; 
 C_{\mu} = \frac{1}{q}[P(r) - 1]
\delta^{\theta}_{\mu} , 
\end{equation}
subjected to the usual boundary conditions for a vortex configuration:
\[
R(0) = 0 \;\; \mbox{and} \;\; P(0) =1 
\]
\begin{equation}
\lim_{r\rightarrow\infty} R(r)=\eta \;\; \mbox{and} 
\lim_{r \rightarrow \infty} P(r) = 0 .
\end{equation}

We also choose a particular form for the pair $(\sigma , A_{\mu})$ responsible 
for the bosonic current along the string and external magnetic field, respectively:
\begin{equation}
\sigma = \sigma(r)e^{i\psi(z)} \;\; \mbox{and} \;\;
A_{\mu} = \frac{1}{e}[A(r)- \frac{\partial{\psi}}{\partial z}]
\delta^{z}_{\mu}
\end{equation}
The pair $(\sigma,A_{\mu})$ is subjected to the 
following boundary conditions
\[
\frac{d\sigma(0)}{dr}=0 \;\; \mbox{and} \;\; A(0)=\frac{dA(0)}{dr}=0 
\]
\begin{equation}
\lim_{r\rightarrow \infty}\sigma (r) =0 \;\; \mbox{and} \;\; 
\lim_{r \rightarrow \infty}A(r) \neq 0.
\end{equation}
With this choice, we can see that $\sigma$ breaks electromagnetism inside the 
string and can form a charged scalar condensate in the string core. Outside the string, 
the $A_{\mu}$-field has a non-vanishing component along the $z$-axis which indicates that
 there will be a non-vanishing energy-momentum tensor in the region exterior to the string. 

With metric given by (8), the Einstein's eqs. read:
\begin{eqnarray}
\beta^{''} & = & 8\pi G\beta e^{2(\gamma - \Psi)} [T^{t}_{t} + T^{r}_{r}] \nonumber \\
(\beta\Psi^{'})^{'} & = & 4\pi G\beta e^{2(\gamma - \Psi)} [T^{t}_{t} + T^{r}_{r}
+T^{\theta}_{\theta} - T^{z}_{z}] \nonumber \\
\beta^{'}\gamma^{'} & = & 
8\pi G \beta e^{2(\gamma - \Psi)}T^{r}_{r} +\beta(\Psi^{'})^{2}
 - \beta(\phi^{'})^{2}  \nonumber \\
(\beta\phi^{'})^{'} & = & 4\pi G \alpha(\phi)\beta e^{2(\gamma - \Psi)} T  ,
\end{eqnarray}
where $(')$ denotes ``derivative with respect to r". The non-vanishing components 
of the energy-momentum tensor (computed using equation (5)) are 
\begin{eqnarray}
T^{t}_{t} &  = & \frac{1}{2}\Omega^{2}(\phi) \{ 
e^{2(\Psi - \gamma)}(R'^{2} + \sigma'^{2}) + \frac{e^{2\Psi}}
{\beta^2}R^2P^2 + e^{-2\Psi}\sigma^2A^2  \nonumber \\
& & + \Omega^{-2}(\phi)e^{-2\gamma}(\frac{A'}{e})^2 + 
\Omega^{-2}(\phi)\frac{e^{2(2\Psi - \gamma)}}{\beta^2}(\frac{P'}
{q})^2 + 2\Omega^2(\phi)V(R,\sigma) \} \nonumber \\ 
T^{r}_{r} & = & -\frac{1}{2}\Omega^2(\phi) \{ e^{2(\Psi - \gamma)}
(R'^2 + \sigma'^2) - \frac{e^{2\Psi}}{\beta^2} R^2P^2 - e^{-2\Psi}\sigma^2A^2 \nonumber \\
& & + \Omega^{-2}(\phi)e^{-2\gamma}(\frac{A'}{e})^2 + \Omega^{-2}(\phi)
\frac{e^{2(2\Psi - \gamma)}}{\beta^2}(\frac{P'}{q})^2 - 2\Omega^2(\phi)
V(R,\sigma) \} \nonumber \\
T^{\theta}_{\theta} & = & \frac{1}{2}\Omega^2(\phi) \{ e^{2(\Psi - \gamma)}(R'^2 + \sigma'^2) 
- \frac{e^{2\Psi}}{\beta^2} R^2P^2 + 
e^{-2\Psi}\sigma^2A^2 \nonumber \\
& & + \Omega^{-2}(\phi)e^{-2\gamma}(\frac{A'}{e})^2 - \Omega^{-2}(\phi) 
\frac{e^{2(2\Psi - \gamma)}}{\beta^2}(\frac{P'}{q})^2 + 2\Omega^2(\phi)
V(R,\sigma) \} \nonumber \\
T^{z}_{z} & = & \frac{1}{2}\Omega^2(\phi) \{ e^{2(\Psi - \gamma)} 
(R'^2 + \sigma^2) + \frac{e^2\Psi}{\beta^2}R^2P^2 - e^{-2\Psi}\sigma^2
A^2 \nonumber \\
& & - \Omega^{-2}(\phi)e^{-2\gamma}(\frac{A'}{e})^2 + \Omega^{-2}(\phi)
\frac{e^{2(2\Psi -\gamma)}}{\beta^2}(\frac{P'}{q})^2 +  2\Omega^2(\phi)
V(R,\sigma) \}
\end{eqnarray}

For the purpose of this analyse, we will divide the spacetime into two regions: an 
exterior region, $r > r_0$, where all fields vanish, except of the magnetic field 
generated by the string; and an interior region, $r \leq r_0$, where all fields contribute
 to the energy-momentum tensor. Conveniently, $r_0$ has about the same order of 
magnitude of the string radius.

\section*{III. Superconducting Cosmic String Solution in Scalar-Tensor Gravity:}

\subsection*{The Exterior Solution:}

In the exterior solution, $r > r_0$, the energy-momentum tensor (15) reduces to:
\begin{equation}
T_{\mu\nu} = \frac{1}{2}F_{rz}F^{rz} diag[g_{tt}, -g_{rr}, g_{\theta\theta}, -g_{zz}]
\end{equation}
We also notice that in this region the trace of the energy-momentum tensor 
is zero: $T=0$. Therefore, the Einstein's equations (14)  can be solved straightforwardly:
\begin{equation}
\beta(r) = Br
\end{equation}
\begin{equation}
\phi(r) = l \ln(r/r_0)
\end{equation}
\begin{equation}
\gamma(r) = m^2\ln (r/r_0)
\end{equation}
\begin{equation}
\Psi(r) = n\ln(r/r_0)- \ln[\frac{(r/r_0)^{2n}+k}{(1+k)}]
\end{equation}
where the constant $n$ is related to $l$ and $m$ through the expression $n^2 = l^2+m^2$. 
We notice that the solutions for the funtions $\gamma(r)$ and $\Psi(r)$ are obtained 
through the equations:
\[
R^t_t = R^{\theta}_{\theta} \rightarrow \gamma''+\frac{1}{r}\gamma' = 0 .
\]
and
\[
R= - 2e^{2(\Psi - \gamma)}(\phi')^2 \rightarrow  \Psi'' +\frac{1}{r}\Psi' -\Psi'^2 = -\frac{n^2}{r^2} .
\]
We point out that in the framework of a scalar-tensorial gravity, the Rainich 
algebra\footnote{L. Witten, ed. {\em ``Gravitation: an Introduction to Current 
Research"} (Wiley, NY, 1962).} is no longer valid. 

The exterior metric is then given by:
\begin{equation}
ds^2 =  (r/r_0)^{-2n} W^2(r) [ (r/r_0)^{2m^2}(dt^2 - dr^2) 
-B^2r^2d\theta^2] - (r/r_0)^{2n}\frac{1}{W^2(r)}dz^2
\end{equation}
where
\[
W(r) = \frac{(r/r_0)^{2n} +k}{1+k}
\]
and the solution for the scalar field is given by eq. (18).

The external current is given by:
\begin{equation}
I \equiv \int \sqrt{-g}J^z dr d\theta =  \frac{2\pi Bn}{1+k}
\sqrt{\frac{k}{G}}
\end{equation}
and the energy density per unit length is:
\begin{equation}
\epsilon(r_*)=\int \sqrt{-g} T^t_t dr d\theta = \frac{Bn}{2G}
\frac{k}{1+k}\frac{(r_*/r_0)^2n - 1}{(r_*/r_0)^{2n}+k}
\end{equation}
The constants $B,l,m,n$ will be determined after the introduction of the matter fields.

\subsection*{The Internal Solution:}
In the internal region all fields contribute to the energy-momentum tensor, 
which makes impossible to solve analytically eqs. (14) with $T^{\mu}_{\nu}$ given in (15). 
Hereafter, we will consider these eqs. in the weak-field approximation. That is, we will
 consider that the exterior and interior metrics are only weakly perturbed from the flat 
space metric. The reason why we can use this approach is twofold:
First, strings of cosmological interest present $G\Omega^2(\phi)M^{t}_{t} \ll 1$, 
where here we have used the source tensor defined by Thorne\footnote{K. S. Thorne, Phys. 
Rev. {\bf 138} (1965), 267.}
\[
M^{\mu}_{\nu} \equiv - 2\pi \int_0^{r} T^{\mu}_{\nu}(r') r' dr'
\]
($M^t_t$ is the energy density per unit length). Second, we can consider that the  
scalar field $\phi$ is a small perturbation on the flat space metric\footnote{M. E. X. Guimar\~aes, 
Class. Quantum Grav. {\bf 14} (1997), 435.}. Hence, 
\[
\phi = \phi_0 + \phi_{(1)} \;\;\; \mbox{and} \;\;\; \Omega(\phi) = \Omega(\phi_0) + 
\Omega'(\phi_0)\phi_{(1)}
\]

Linearising eqs. (14) to the lowest order in $G\Omega^2(\phi)M^{\mu}_{\nu}$ we have the solutions:
\[
\frac{d\beta}{dr} \sim 1 - 8\pi G\Omega^2(\phi_0)\int_0^r [\sigma^2 A62 + R^2P^2 + 
2\Omega^2(\phi_0)V(R,\sigma)] r dr
\]
In $r=r_0$, we have:
\begin{equation}
\frac{d\beta(r_0)}{dr} = B \sim 1 - 8\pi G\Omega^2(\phi_0)\int_0^{r_0} [\sigma^2 A62 + R^2P^2 + 
2\Omega^2(\phi_0)V] r dr
\end{equation}
\[
r\frac{d\Psi}{dr} \sim -8\pi G\Omega^2(\phi_0)\int_0^r [\sigma^2 A^2 + 
\Omega^{-2}(\phi_0)(\frac{A'}{e})^2 - \Omega^{-2}(\phi_0) (\frac{P'}{rq})^2 + 
\Omega^2(\phi_0)V(R,\sigma)] r dr
\]
In $r=r_0$, we have:
\begin{eqnarray*}
\frac{d\Psi(r_0)}{dr}  & \sim &  -\frac{8\pi  G\Omega^2(\phi_0)}{r}\int_0^{r_0} 
[\sigma^2 A^2 + \Omega^{-2}(\phi_0)(\frac{A'}{e})^2 - \Omega^{-2}(\phi_0) (\frac{P'}{rq})^2 + 
\Omega^2(\phi_0)V(R,\sigma)] r dr \nonumber \\
& & - 8\pi G\int_{r_0}^{r} (\frac{A'}{e})^2  r dr
\end{eqnarray*}

The second integral in the r.h.s. of eq. above is just the expression of the energy density 
(23). On the other hand, linearising the exterior metric function $\Psi(r)$ given in (20), we have:
\begin{equation}
n\frac{1-k}{1+k} \sim 8\pi G \Omega^2(\phi_0) \int_0^{r_0}
[\sigma^2 A^2 + \Omega^{-2}(\phi_0)(\frac{A'}{e})^2 - \Omega^{-2}(\phi_0) (\frac{P'}{rq})^2 + 
\Omega^2(\phi_0)V(R,\sigma)] r dr
\end{equation}

Linearising the eq. for the scalar field, 
\[
\frac{d\phi}{dr} \sim \frac{4\pi G}{r} \alpha(\phi_0)\Omega^2(\phi_0)\int_0^{r} T(r') r' dr' .
\]
Making the junction with the exterior solution in (18), we have:
\begin{equation}
l \sim 4\pi G \alpha(\phi_0)\Omega^2(\phi_0) \int_0^{r_0} [R'^2 + \sigma'^2 + R^2P^2 + 
\sigma^2A^2 + 2\Omega^2(\phi_0)V(R,\sigma)] r dr
\end{equation}

\subsection*{Bending of Light Rays:}

A light ray coming from infinity in the transverse plane has its trajectory deflected, 
for an observer at infinity,  by an angle given by:

\[
\Delta\theta = 2 \int_{r_{min}}^{\infty} dr [-\frac{g^2_{\theta\theta}l^{-2}}{g_{rr}g_{tt}} 
- \frac{g_{\theta\theta}}{g_{rr}}]^{-1/2} \,\, - \pi 
\]
where $r_{min}$ is the distance of closest approach, given by 
$\frac{dr}{d\theta} =0$:

\[
\frac{g_{\theta\theta}(r_{min})}{g_{tt}(r_{min})} = -l^2
\]

which gives in turn:

\[
\frac{r_{min}}{r_0} = (\frac{l}{Br_0})^{1/(1-m^2)}.
\]

We can now evaluate the deficit angle to first order in $G\Omega^2(\phi_0)$. 
Performing an expansion to linear order in this factor, in much the same way as 
Peter and Puy\footnote{P. Peter and D. Puy, {\em Phys. Rev. D} {\bf 48} (1993), 5546.} we find:

\begin{equation}
\Delta\theta = \frac{2}{B(1-m^2)}[\frac{\pi}{2}(1+m^2\ln\frac{l}{Br_0})-m^2\nu]
-\pi ,
\end{equation}
where we have defined the quantity $\nu$ as 
\[
\nu \equiv - \int_0^1\frac{\ln s}{\sqrt{1-s^2}} ds \sim 1.08879
\]
with $s \equiv \frac{l}{Br_0}(\frac{r}{r_0})^{m^2 -1}$. 

\section*{IV. Discussions and Conclusion:}

In this work we have considered a superconducting vortex arising from 
a $U(1) \times U'(1)$ model coupled to a scalar-tensorial gravity. 
We found the metric in the weak-field approximation. As expected, 
the dilaton contributes to the observable quantities (e.g., the deficit angle), 
albeit in this approximation its contribution is very small. In particular, 
the deficit angle given by expression (27) depends explicitly on the 
current and the dilaton, although eq. (27) is very difficult to be analysed. 
In summary, the results found in the present work deserve further 
numerical analysis, which is being done 
currently\footnote{C.N. Ferreira, M.E.X.G. and J.A. Helay\"el-Neto, 
{\em in preparation}.}.

\end{document}